\documentclass[mpicover]{mpireport-inf}
\usepackage{times}
\usepackage{amsmath,epsfig}
\usepackage{subfigure}
\usepackage{multirow}
\usepackage{subfigure}
\usepackage{boxedminipage}
\usepackage{graphics}
\usepackage{algorithm, algorithmic}
\usepackage{amsmath,amssymb,amsfonts}
\usepackage{graphicx}
\usepackage{latexsym}
\usepackage{color}
\usepackage{url}
\usepackage{appendix}

\newcommand{\comment}[1]{}
\newcommand{\eat}[1]{}
\newcommand{\Tag}[1]{{\tt #1}}

\title{Generating Concise and Readable Summaries of XML Documents}
\author{Maya Ramanath,
Kondreddi Sarath Kumar,
Georgiana Ifrim}

\repnumber{5-002}
\setrepyear{2009}
\setyear{2009}
\setmonth{May}

\begin{addresses}
Maya Ramanath\\
Max-Planck Institute for Informatics\\
66123 Saarbr\"ucken, Germany\bigskip\\
Kondreddi Sarath Kumar\\
Max-Planck Institute for Informatics\\
66123 Saarbr\"ucken, Germany\bigskip\\
Georgiana Ifrim\\
Bioinformatics Research Center\\
Aarhus University\\
DK-8000 Aarhus C\\
\end{addresses}


\begin{document}
\begin{abstract}

XML has become the de-facto standard for data representation and exchange, resulting 
in large scale repositories and warehouses of XML data.
In order for users to understand and explore these large collections, a summarized, bird's 
eye view of the available data is a necessity. In this paper, we are interested 
in \emph{semantic} XML document summaries which present the 
``important'' information available in an XML document to the user. In the 
best case, such a summary is a concise replacement for the original 
document itself. At the other extreme, it should at least help the user make an 
informed choice as to the relevance of the document to his needs. In this
paper, we address the two main issues which arise in producing
such meaningful and concise summaries: i) which tags or
text units are important and should be included in the summary, ii) how to generate summaries of different sizes.
 We conduct user studies with different real-life datasets and show that our methods are useful and effective in practice.
\end{abstract}

\makempicover
\tableofcontents

\chapter{Introduction}

With the ubiquity of XML as the format of storage and exchange of data, we can 
expect to see ever-growing repositories of XML documents. Exploration of these 
collections requires the use of a diverse set of 
tools ranging from classifiers, clustering tools, data visualizers to mining 
software. One of the ways in which \emph{human-centric} exploration can be made easier is to 
provide the user with a concise, summarized view of the information 
contained in an individual or in a set of documents. Consider the following scenario. Suppose there is a large corpus of XML documents, each of which describes a movie released in the last 30 years (for example, extracted from IMDB). A movie enthusiast wants to make a list of interesting movies based on various criteria, such as, the genre, lead actors, directors, etc. She first decides to narrow the focus to just thrillers. However, she then has to look into each document individually, since only then is it possible for her to tell whether the combination of actors, directors, etc. interests her. This would be time-consuming if the documents in question contain hundreds of tags each. Instead, if short summaries of each document could be presented to her, she could use these summaries to filter out movies she certainly would not be interested in. The generation of such summaries is the problem we address in this paper.

A \emph{generic summary} summarizes the entire contents of the document by identifying and possibly rewriting the most important content. The implicit assumption regarding the user's information need is that she is interested in knowing ``what is in the document'' without having to read the document in its entirety. In this paper we propose techniques to \emph{automatically} generate concise, readable summaries of XML documents subject to size constraints.%

\begin{figure*}[ht]
	\begin{center}
	\subfigure[Original Document]{
		\includegraphics[scale=0.45]{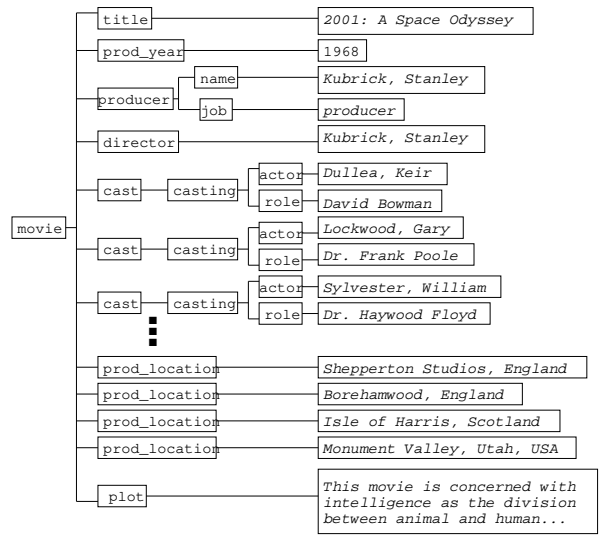}
		\label{fig:example-orig}
		}
		\hspace*{0.15in}
	\subfigure[Summary]{
		\includegraphics[scale=0.45]{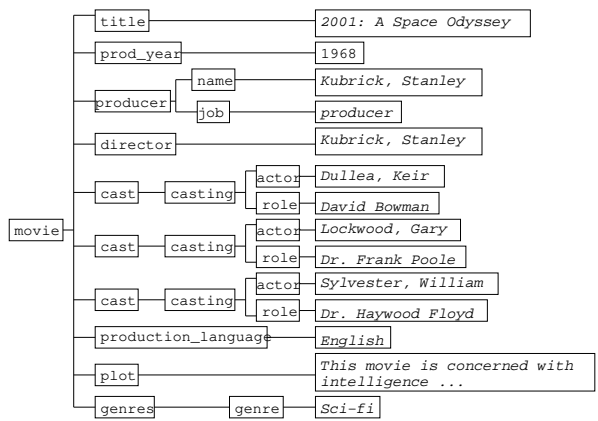}
		\label{fig:example-des}
		}
	\caption{``2001: A Space Odyssey'' -- Snippet of original document and a possible summary}
	\label{fig:intro-example}
	\end{center}
\end{figure*}

As a concrete example of the type of summaries we are interested in, consider 
Figure \ref{fig:intro-example}. The original document in Figure \ref{fig:example-orig} describes the  movie ``2001: A Space Odyssey'' (from IMDB). All the tags in this snippet have 
semantics associated with them (they are not tags for formatting or display), 
and there are short pieces of information at the leaf level, such as the title 
of the movie, its director, genres, etc. A concise summary of this snippet is 
shown in Figure \ref{fig:example-des} where only the ``important'' information has been retained -- some tags and their text values have been dropped completely (several {\tt actor}s and {\tt production\_location}s for example). The resulting summary is 
shorter but conveys all the important information from the original
document.

\section{Challenges in XML Summarization}

Informally, a summary is useful if, at the very least, it helps the user decide whether a
particular document is worth looking into in its entirety or not. The best
summary would encapsulate most or all the salient points of the
document and could, in many cases, serve as a replacement for the original
document. However, generating such a summary may often involve a trade-off between \emph{coverage} and \emph{importance}. Important content may have to be sacrificed in order to improve coverage. And coverage may have to be reduced to ensure all important content is included. The right balance between the two is required, and this balance has to be achieved  given a limit on the summary size. While importance and coverage are factors that need to be considered in both text and XML summarization, there is one additional source of complexity in XML summarization. That is, importance and coverage have to extend to both structure as well as content. And since structure and content play different roles in the document and have different characteristics, different techniques may be needed to deal with them.

\section{Our Approach}

We regard the problem of generating XML summaries as a two-stage problem. First, we separately rank tags and text according to a notion of their importance. Next, we construct the summary based on the tag and text scores. The choice of tag-text pairs is made such that the summary reflects only their relative importance in the document, thus achieving a balance between including importance and coverage.

\subsection{Contributions and Organization}
\noindent
Our contributions are the following:
\begin{itemize}

\item We present a formal model for the generation of XML summaries.

\item We propose techniques for ranking tags and text values based on their importance in the document and the corpus.

\item We propose an algorithm which takes the ranked tags and text values and constructs a summary while strictly adhering to a size limit.

\item Finally, we test our techniques for summary generation with a user study using real-life datasets.
\end{itemize}

The rest of this paper is organized as follows. Related work is discussed in Section \ref{sec:2relwork}. Section \ref{sec:3model} discusses the
data characteristics and our model for summarization. Section \ref{sec:4ranking} discusses various techniques to prioritize tags and text values to be included in the summary. Section \ref{sec:5sumGen} describes the algorithm for generating a summary given a size limit. Section \ref{sec:6expts} reports on our user study.  Finally, we conclude in Section \ref{sec:7conc}.

\chapter{Related Work}
\label{sec:2relwork}

Text summarization is a well-developed field (see, for example, \cite{Hahn:IEEEComputer2000} for an overview), dedicated to developing techniques for 
summarizing text documents. In a nutshell, the aim is to present \emph{important} and \emph{non-redundant} information contained in a document to the user in a concise and readable manner. One way to tackle the problem of summarization is to look at it as a ranking problem -- text spans (say, sentences) are ranked according to a certain set of features and only the top-ranked spans are included in the summary. We follow this approach in our work. That is, we rank both tags and text values and consider them for inclusion in the summary according to their rank. However, the techniques for text summarization are not directly applicable in our context for two reasons: i) the structure of an XML document may be as important as the text (the tags and tag hierarchy in XML are meant for providing additional semantics) ii) the textual values which appear in XML documents are not always free-flowing text, for which text summarizers are most suitable.

XML has been used as markup in text documents to make them more feature rich and enable better text summarization (see for example, \cite{Litkowski:DUC2005} for an overview). Summarization techniques are described in \cite{Amini:JIR2007} for such feature-rich XML documents. However, the goal of these techniques is still to rank and extract the best sentences to be included in a summary. And the structural (XML) features of the document are specifically made use of to improve upon previous techniques for summarizing text spans. Hence, the kind of XML document that is being summarized is still predominantly text while certain parts of the text are tagged. In contrast, our work deals with documents which may or may not have free-flowing text.

The work presented in \cite{Yu:VLDB2006} deals with XML Schema summarization, where the goal is to present important schema elements to make large XML \emph{schemas} easily readable by the user. Our work, on the other hand, is interested in summarizing XML \emph{documents} of which structure is only one part and the data the other. AxPRE summaries, proposed in \cite{Consens:ICDE2008} generates summaries of the structure of a corpus of XML documents. The user can then \emph{interactively} explore the repository by selectively expanding parts of the summary. We differ from this work in two ways - first, we do single document summaries and second, we take into account the text values present in the document. In contrast, AxPRE can be regard as a multi-document structural summarizer.

Other approaches to representing XML data in a concise manner include compression (for example, \cite{Ferragina:WWW2006}) and statistical summaries (for example, \cite{Freire:SIGMOD2002}). However, the many tools for compression focus on efficient query processing and not on the readability for the end-user. And statistical summaries are used mainly for cardinality estimation which feed into the query optimizer. Work on building tools for database exploration (for example, \cite{Laks:SIGMOD2003, Saint-Paul:VLDB2005}) is also relevant in our context. However, our work looks at \emph{data-oriented XML} while the focus of these tools is on developing techniques to summarize the entire database (of relational tuples). On the other hand, a document-oriented view is taken, for example, in \cite{Dakka:ICDE2008}. But, there the aim is to extract \emph{facets} from a database consisting of \emph{text} documents while we aim at presenting the user with a concise and readable summary of individual \emph{XML} documents.

Generating snippets of XML query results \cite{Huang:sigmod2008} is close to our work, but our setting is different in that we consider stand-alone XML documents and rank elements without any query bias. Finally, our own previous work \cite{Ramanath:dbrank2008} presented ideas for generic XML document summarization and a user study illustrating its effectiveness. The current paper provides a summarization model and formalizes the scoring functions. In addition, we present a more extensive user evaluation and analysis of the results.

\chapter{Summarization Model}
\label{sec:3model}

\section{A First Attempt}

One appealing scenario for generating summaries is to represent a document $D$ as a set of its tag-text pairs $s_{ij}$, thus $D=\{s_{ij}|i \in \{1 \dots |Tags|\}$ and  $j \in \{1 \dots |Texts \; in \; Tag_i|\}\}$. Let $D$ be a document such that $|D| = n$. Let $S$, the summary of $D$, have size $m$, i.e. it contains $m$ tag-text pairs selected from $D$. The potential number of candidate summaries for $D$ is then $n \choose m$ which makes it prohibitively expensive to generate each summary, score it and return the top-ranked summary. A simple alternative would be to estimate the $m$ most likely tag-text pairs in $D$, and return that as its summary $S$. This method, however does not work well, as shown in the following example.

\begin{table}
\small
\begin{center}
\begin{tabular}{|c|c|c|c|c|}
\hline
{\bf Tags} 	&	{\bf Total} & {\bf Prob. of} & {\bf Prob. of }	& {\bf Joint prob.} \\
          	& {\bf \#tags} &  {\bf Tags}	   & {\bf Text given}	& 										\\
          	&	{\bf in doc} 	& &             {\bf  Tag}&\\
\hline
\Tag{title} &		1               & 0.1          & 1									 & 0.1\\																		
\hline
\Tag{actor}   & 4               & 0.4       		& 0.25, 0.25 $\dots$ 		& 0.1, 0.1, $\dots$ \\
\hline	
\Tag{keyword} & 3 				      & 0.3       		& 0.33, 0.33 $\dots$ 		&	0.1, 0.1 $\dots$ \\
\hline
\Tag{trivia}  & 2 		        	& 0.2       			& 0.5, 0.5 							&  0.1, 0.1\\
\hline 
\end{tabular}
\caption{Probability distribution on tag-text pairs in document with 10 elements.}
\label{tab:jointprob}
\end{center}
\end{table}

\normalsize

Let document $D$, chosen from a movie corpus, have the distribution shown in Table \ref{tab:jointprob} on its tags and their corresponding text. Without a proper scoring mechanism for text values (all text values have equal probability, given the tag), we note that: i) simply computing the joint probability tells us absolutely nothing about the relative importance of the tag-text pairs, ii) we would have to rank \Tag{actor}, \Tag{keyword} and \Tag{trivia} above \Tag{title}, even though the movie title is probably the ``must-have'' tag in any movie summary. This is the direct consequence of computing the probability based on the ``local'' frequency of occurrence -- that is, the number of times a tag occurs in a document may not directly correspond to its importance, and finally, iii) choosing the summary with the maximum likelihood, of say, size 4 elements, would choose only \Tag{actor}s, thus completely ignoring coverage.

\section{Our Approach}

In order to address the above problems, we start by first defining scoring functions for both tags and text values separately. Since clearly, the frequency of occurrence of a tag in the document does not correlate with its importance, our scoring functions are based on a closer examination of the role of tags in XML and includes the corpus statistics. Second, we provide methods to score text values which occur under the same tag. This is based on the premise that we can only compare apples with apples -- that is, it makes more sense to compare an actor with another actor (for example, ``Kate Winslet'' with ``Billy Zane'' in Titanic) and say which of them is more important (``Kate Winslet''), than comparing an actor with a keyword (``Kate Winslet'' with ``Iceberg''). Hence, our text ranking is ``local'' (within the tag context), while our tag ranking is ``global'' (within the entire document). The scoring functions for both take into account both the document as well as the corpus statistics. Finally, we note from the previous discussion that choosing a summary which maximizes likelihood does not ensure coverage. Instead, we approach the summary generation problem as a two step process: first, we constrain the structure of the summary based on an \emph{importance distribution} inferred from the structure of the document and the corpus; second, given the fixed structure we can focus on selecting the most important text associated with that structure.

\subsection{Summarization Framework}

\begin{figure}[t]
\begin{center}
\includegraphics[scale=0.65]{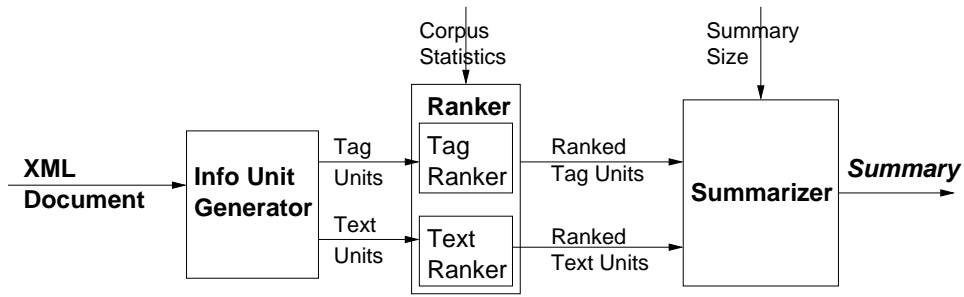}
\caption{Steps in Summarization}
\label{fig:summary-arch}
\end{center}
\end{figure}

The various components involved in our summarization framework are shown in Figure \ref{fig:summary-arch}.
The XML Document is taken as input into an \emph{Information Unit Generator} module. This module generates two types of information units -- tag information units and text information units. Following text summarization techniques, these sets of information units are ranked according to importance by the \emph{Ranker} module which also takes the corpus statistics as input to its ranking functions. The \emph{Summarizer} module takes as input the ranked lists of tag and text information units, along with the size constraint. It chooses tag-text pairs to be included and rewrites them appropriately (for example, to reflect document order) to produce the final summary.

Next, we explain the functionality of these components in more detail.

\subsection*{Information Unit Generation}

An XML document has two different types of content -- tags and text.
They play distinct roles. Tags can be regarded as the metadata for a set of
documents -- that is, tags, nesting of tags and their value types are
defined to express a specific class of information (for example, {\it movies}). This information could be encoded into schemas or DTDs which
are typically much smaller than the corpus data. On the other hand,
text values are required to ``instantiate'' a specific
document. Hence, the statistical properties of tags and text differ considerably. Unlike tags which are highly redundant in the context of a corpus, text values are much less so. For this reason, we need to use different techniques to identify important tags and important text values. Our first step toward this goal is to generate separate sets of ``information units'' for tags and text from the document. We then rank each set of information units using different scoring functions.

\begin{figure}[t]
\begin{center}
		\includegraphics[scale=0.65]{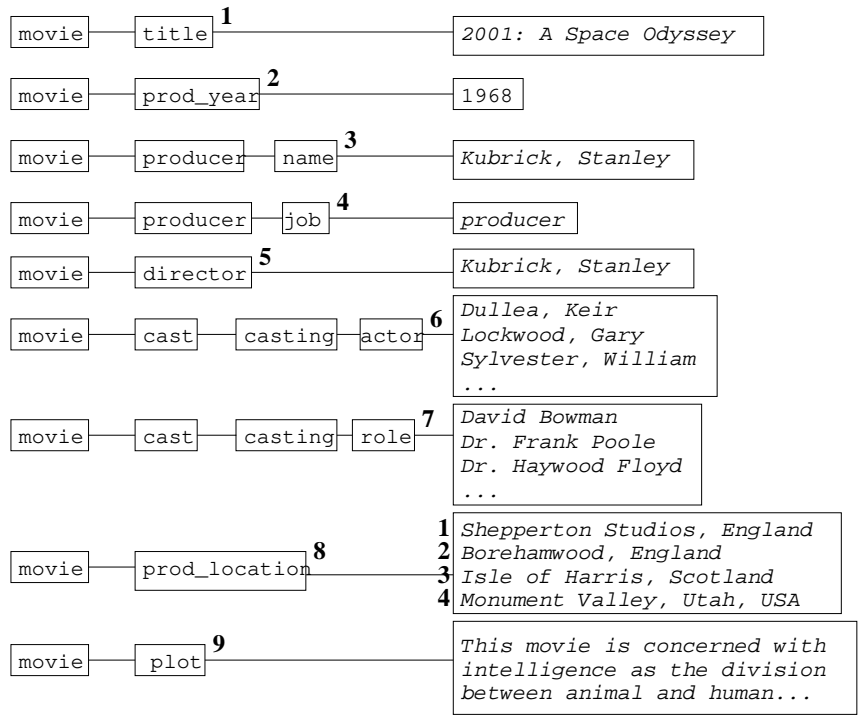}
	\caption{Example document decomposed into its tag and text information units}
	\label{fig:tag-text}
\end{center}
\end{figure}

Figure \ref{fig:tag-text} shows how the movie ``2001: A Space Odyssey'' is decomposed. First, the tag information units are identified -- that is, each unique path from the root to a leaf (without the text value). Second, the text information units are constructed by putting together text values corresponding to {\bf each tag information unit}. Thus text information units are always associated with a tag context. The ranking of tag information units gives a global ordering on tags, while the ranking of text units gives a local ordering, within a tag context.

Figure \ref{fig:tag-text} shows how the tag units (on the left of the figure) and the text units (on the right of the figure) are related to each other. When no distinction between tag and text units is required, they are together termed \emph{information units}.

\subsection*{Ranker}
The ranker module takes the tag and text information units as input and scores them according to an importance measure. Corpus statistics form a crucial input to the scoring functions. A document by itself may not give us enough information about the importance of tags or text units, as shown in the previous section. Instead, a measure of how tags and text values are distributed in a large corpus containing similar information gives us insights into what could be considered important. Note that in the ideal case, we would have an expert who would study the corpus and tell us how to measure the importance of tags and text units. However, this is neither scalable nor often practical. Hence our aim is to define some general principles which would work for many different kinds of documents and corpora. And a crucial part of these principles is the use of corpus statistics. The scoring mechanisms are described in more detail in the next section.

\subsubsection*{Summary Generation} 

Using the ranker module, we basically estimate a distribution of importance on tags and text units, from the document and the background corpus. Our summary generation module takes these distributions as input in order to generate the summary of the required size. As previously mentioned, the summary is a sample of the document based on the importance distribution computed by the ranker module. In addition to this sample, the summary generation unit may also rewrite the summary to make it more readable. We currently support only one kind of rewrite function: the order of tags and text values in the summary will reflect the document order (and siblings in the summary will also be siblings in the original document). However, more complex rewritings are possible and are briefly discussed in Section \ref{sec:7conc}.

In the next section we provide details on our design of the scoring functions for both tags and text units. We then use the ranked lists to construct the summary and describe the process in Section \ref{sec:5sumGen}.

\chapter{Ranking Model}
\label{sec:4ranking}

\section{Ranking Tag Units}
\label{sec:ranking-tags}

\noindent
We consider 2 criteria for considering a tag $t$ important:

\noindent
{\bf Typicality:} If $t$ is salient in the corpus, then it is very likely that it defines the context of the documents. For example, \Tag{title} is the most salient tag in the corpus, since it is present in all documents. And clearly, it sets the context for the rest of the document -- that the given document is about the movie with the given title.

\noindent
{\bf Specialty:} If $t$ is more or less frequent than in the ``average'' document in the corpus, then it is likely to denote a special aspect of the current document. For example, \Tag{production\_location} may occur once typically, but if the current document has 10 of those, then it implies that the film was shot in an unusually large number of locations. Also, if the current document contains \Tag{oscar\_winner} while the average document does not, then that too should be considered special.

Our scoring function is a mixture model of two components (typicality and specialty) with a parameter $\alpha$ controlling the influence of each and is defined as follows.

\begin{equation}
\label{eqn:score}
P(T_i) = \alpha P_{typ}(T_i) + (1-\alpha) P_{spe}(T_i)
\end{equation}

where $P(T_i)$ is the probability of choosing tag $T_i$, $P_{typ}(T_i)$ and $P_{spe}(T_i)$ are the probabilities of choosing $T_i$ based on its typicality and specialty respectively, the parameter $\alpha$, $0 \leq \alpha \leq 1$ is set by the user or learned through examples. It now remains for us to describe how to compute the probabilities $P_{typ}$ and $P_{spe}$.

\paragraph*{Typicality}

As mentioned before, the typicality of the tag unit refers to
``common knowledge'' in the corpus. If it occurs in most or all documents,
then the tag unit is considered very typical and ranked high. We quantify the
typicality of a tag unit by measuring the fraction of documents in
which the tag unit occurs (document frequency). That is, we define the typicality of tag unit $T_i$ as,
\begin{equation}
typ(T_i) = \dfrac{|D | T_i \in D|}{|D|} \nonumber
\end{equation}
where the numerator is the document frequency of $T_i$ (number of
documents in which $T_i$ occurs) and the denominator is the total
number of documents in the corpus $C$.

The tags can now be ranked in order of their typicality values -- the
higher the typicality, the higher the rank. We normalize the typicality of tags to get a probability distribution on the typicality values as follows:

\begin{equation}
P_{typ}(T_i) = \dfrac{typ(T_i)}{\displaystyle\Sigma_j{typ(T_j)}}
\end{equation}

\paragraph*{Specialty}

The specialty of a tag is characterized by how different the frequency of the tag in the current document is from an average document in the corpus. The current document could contain a larger number or a smaller number of instances of a particular tag than the average document.

In order to construct the average document, we simply estimate for each tag, its average number of occurrences per document in the corpus:

\begin{equation}
count_{avg}(T_i|C) = \dfrac{|T_i|T_i \in C|}{|D|} \nonumber
\end{equation}

\noindent
where the numerator is the number of times $T_i$ occurs in the corpus and the denominator is the number of documents in the corpus. Now, in order to compute how much a tag $T_i$ deviates from this average document we use the following:

\begin{equation}
dev(T_i) = \max{\left \{\dfrac{|T_i|T_i \in D|}{count_{avg}(T_i|C)}, \dfrac{count_{avg} (T_i|C)}{|T_i|T_i \in D|} \right\}} \nonumber
\end{equation}

\noindent
where $dev(T_i)$ is the maximum of the ratio between the number of tags in the current document and the number of tags in the average document and its reciprocal. We use the maximum value so that tags which occur less number of times as well as those which occur a greater number times than the average document are given equal consideration (for example, an unusually large number of \Tag{production\_location}s must be as important as an unusually small \Tag{cast} compared to the average document). 

We then compute the specialty as follows, where the numerator denotes the deviation of tag $T_i$ and the denominator is the normalizing factor to get a probability distribution of the specialty of tags.
\begin{equation}
P_{spe}(T_i) = \dfrac{dev(T_i)}{\Sigma_j dev(T_j)}
\end{equation}

\section{Ranking Text Units}
\label{sec:ranking-text}

The problem of ranking text units is more complex than that of
tags. This is mainly because of the many different forms of text that
can occur in a document. For example, a document could contain
free-flowing long text values, short text values, entities, etc. For
each of these kinds of text, a different ranking mechanism would make
sense. We divide text into the following categories: i) entities and
ii) regular text. Entities are treated holistically and our system
currently supports proper names. Regular text could be long or short
and can be reduced to a set of terms. In addition to these two types
of text, we make another distinction with respect to their
occurrence. Ideally, we should rank text values of a given text unit
with respect to the other text values in the same unit. However, this
is possible only if the terms (or the entity) occur multiple times in
the text unit. When such redundancy is not to be found within the context
of the tag unit, we need to change the context to take into
consideration the document and the corpus. Examples of such text units
could include the list of actors in a movie or the genres of the
movie, etc. Examples of text units which are redundant within the
context of their tag unit include trivia items, plots, goofs etc.

Our general model for text units, regardless of whether or not they
have redundancy in the tag, document or corpus context is a mixture
model defined as follows:

\begin{equation}
P(t_j|D,T_i) = \lambda P(t_j|D,c(T_i)) + \mu P(t_j|D) + (1-\mu-\lambda)P(t_j|C)
\end{equation}

where the first term, $P(t_j|D,c(T_i))$ denotes the probability of
choosing text value $t_j$ within the \emph{context} of $T_i$ (denoted
$c(T_i)$). The second and third terms, $P(t_j|D)$ and $P(t_j|C)$
denote the probability of $t_j$ in the document and the corpus
respectively.

The probability $P(t_j|C)$ mainly comes into play when $t_j$ has
little or no redundancy in the tag context $T_i$\footnote{Note that
without any redundancy, the best conclusion we can come to is that
each value is equally important}. In these cases $\lambda$ and $\mu$ have to be set empirically or
learned through examples. These values can be tuned depending
on the corpus.

We next discuss how to estimate each of the above three probability distributions. Note that it is fairly easy to determine whether or not a text unit is redundant within its tag unit by examining the document.

\subsection{Text with Redundancy in its Tag Context}

We would like to choose a set of text values which are representative
of the text unit, while also being as diverse as possible. That is, we
should choose values which are important while simultaneously
increasing coverage.  For the first goal of extracting the most
representative or important of the text values, we use the centroid
query method. For the second goal of ensuring diversity, we utilize
the concept of maximal marginal relevance (MMR) proposed in
\cite{Carbonell:sigir1998}.

\paragraph*{Centroid query method} Let TEXT $= \{t_i\}$, $1 \leq i \leq n$ be the text unit for the tag
$T$. Let, TERM $= \{trm_i | trm_i \in t_j, 1 \leq j \leq n\}$ be the
set of terms occurring in any of the $t_i$'s. Let $F = (trm^i)$ be the
sequence of terms from TERM sorted by their frequencies of
occurrence in TEXT, where $i$ denotes the rank of $trm$. We now
choose the top $m$ terms from $F$ to be the centroid query $Q$. That
is, $Q = \{trm^i| 1 \leq i \leq m\}$. The set $Q$ contains terms which
are representative of the text unit. We now compute the relevance (or
similarity) of each text unit $t_i$ with respect to $Q$:

\begin{equation}
R(t_i) = {\displaystyle\sum_{1 \leq j \leq m}{\dfrac{freq (q_j | q_j \in t_i)}{\displaystyle\sum_k count(t_k | q_j \in t_k)}}}
\end{equation}

where the numerator is the term frequency of $q_j$ in text value $t_i$
while the denominator is the number of text values $t_j$ which contain
the term $q_j$. The final score is the sum over all terms of
$Q$\footnote{This is analogous to the $tf.idf$ scoring.}.

\paragraph*{Diversity} The above relevance gives us a ranking of text values from the most
relevant to the least relevant. However, as stated before, our aim is to
increase diversity, while simply using the ranking above would
give us values which are ``more of the same''. In order to increase
diversity, we use the MMR metric proposed in
\cite{Carbonell:sigir1998}. The idea of the MMR metric is to do a
re-ranking of the text units once a particular text unit has been
included in the summary. The re-ranking considers the text units not yet
included in the summary and calculates a new ranking for these text
units based on their similarity to the already included text units and
their relevance rank. In order to calculate the similarity between two
text values, we first eliminate stop words in both text values and
stem all the terms. The number of common terms between the two values
gives us an estimate of their similarity.

Let $T = \{t^1,t^2,...,t^m\}$ be the set of text values already
included and let $T' = \{t_1,t_2,...,t_k\}$ be the set of text values
yet to be included. To compute a new score for the elements of $T'$,
we use the following formula:

\begin{equation}
S(t_i) = \beta R(t_i) - (1-\beta)\displaystyle\max_{t^j \in T}(sim(t_i, t^j))
\end{equation}

where $R(t_i)$ is calculated as shown above and $sim (x,y)$ is
calculated as \\ $|terms(x) \cap terms(y)|$ where $terms(x)$ and
$terms(y)$ are the set of terms in text values $x$ and $y$
respectively.

In order to get a final ranked list of text values, we need to repeat
the process $n-1$ times. That is, we first choose the highest ranked
text value according to $R(t_i)$. Then compute $S(.)$ for the
remaining text values to choose the second text value. Similarly, we
repeat to choose the third text value and so on until we get a final
ranked list. Let $T = (t^1,t^2,...,t^n)$ be the $n$ text values in
text unit $T$ in ranked order. If there are negative scores in $T$, we
perform the normalizing step of adding the minimum score in $T$ plus
$1$ to all scores to convert them into positive values. We can then
define,

\begin{equation}
P(t_i|D,c(T_j)) = \dfrac{S(t_i)}{\displaystyle\sum_j S(t_j)}
\end{equation}

\subsection{Text with Almost no Redundancy at Tag Level}

When there is no redundancy of text values at the tag level, then we
need to look at the document and possibly the corpus in order to rank
them. The document-context probabilities are calculated as,

\begin{equation}
P(t_i|D) = \dfrac{|t_i|t_i \in D|}{\displaystyle\sum_k |t_k|t_k \in D|}
\end{equation}

where the numerator is the number of occurrence of $t_i$ in the
document and the denominator is the sum total of occurrences of
\emph{all} text values in this text information unit.

Analogously, at the corpus level, we simply count the number of occurrences of $t_i$ and normalize it as above to get $P(t_i|C)$.

\section{Handling Co-occurring Tags (and Text Values)}

So far, we have described the ranking of tag units assuming that they
occur independently of one another. However, since XML has a tree
structure, it is often the case that we find \emph{related} tag units
-- tag units which are siblings of one another. An example of such an
occurrence is a \Tag{role} occurring along with \Tag{actor}. Clearly, the text values corresponding to these two tags should co-occur in the summary. For example, we would like a role to appear with the actor who played that role, rather than simply pair up the top-ranked actor value with the top-ranked role value. One further aspect to consider is whether it makes sense to include all co-occurring tag units in the summary or only a subset of them is still acceptable. For example, it is perfectly acceptable for an \Tag{actor} to appear without the corresponding \Tag{role} that he played, but would likely make no sense when a \Tag{role} appears without the corresponding \Tag{actor}.

We address these issues with our rankings of tag and text
units as follows. Let the co-occurring siblings under
consideration for inclusion in the summary be: $T_{sib} =
(T_1,T_2,...,T_k)$. Let $rank(T_1) \geq rank(T_2) \geq...\geq
rank(T_k)$. Then,

{\bf Case 1:} If $rank(T_1) = rank(T_2) = ... = rank(T_j)$, $j
\leq k$, then \emph{all} of $\{T_1,...,T_j\}$ should be included in
the summary at one shot. Moreover, for the inclusion of text values
corresponding to these siblings, we choose a $T_i$ at random from
among $\{T_1,...,T_j\}$ and include its best ranked text value
$t_i$. Then the \emph{corresponding} $t$s of the remaining tag units,
\emph{regardless} of their rank are chosen for inclusion. If the desired size of the summary is exceeded because of this inclusion, then we can decide to either include  only a subset and satisfy the size requirements or to exceed the size limit. We currently consider size to be a hard constraint and only include a subset.

{\bf Case 2:} If $rank(T_1) > rank(T_2)$ (implying that it is
also greater than the rest of the $T_i$s), then only $T_1$ is included
in the summary at the current time along with its best ranked text
value $t_1$. At a later stage in the summary construction, if we also
include $T_2$, then we include $T_2$ as a sibling of $T_1$ and choose
a text value which co-occurs with $t_1$. The principle is repeated for
the rest of the siblings. The reasoning is that because of its higher
rank, $T_1$ plays a more dominant role among the siblings and can
occur by itself (which indeed it has in the document and/or corpus,
otherwise, it would not have a higher score than the others).

We have now described techniques to rank tag and text units. The next
step is to construct a summary of the required size, given these
rankings.

\chapter{Generating the Summary}
\label{sec:5sumGen}

We discussed the ranking of tags and text values and the rationale for generating a summary $S$ by the process of sampling its structure and its content 
from the corresponding ``importance'' distributions for tag and text units.
In this section, we discuss summary generation in detail and outline practical issues that arise during this process and our solutions to overcome them.

\begin{table}[ht]
\begin{center}
\begin{tabular}{|c|c|c|}
\hline
{\bf Tag} & {\bf Prob.} & No. of tags\\
          &             & in summary\\
\hline
\Tag{actor}   & 0.5       & 15\\
\hline
\Tag{keyword} & 0.3       & 9\\
\hline
\Tag{trivia}  & 0.2       & 6 \\
\hline 
\end{tabular}
\caption{Number of tag units in a summary of size 30 spans}
\label{tab:rank}
\end{center}
\end{table}

\vspace*{-0.1in}
As a first step, we compute the number of tags of each type which should occur in the summary based on the estimated distribution on tags presented in Section \ref{sec:4ranking}. To give a simple example, suppose we want a summary of 30 spans of a document containing just 3 tags -- {\tt actor}, {\tt keyword} and {\tt trivia}. Let their probabilities be as shown in Table \ref{tab:rank}. Hence, the summary should contain 15 {\tt actor}s, 9 {\tt keyword}s and {\tt 6 trivia} items. Once the structure of the summary is fixed, we select the most likely text units for each of the tag types, in order to build the necessary spans.

However, we encounter a first problem in this setting. The document may not contain the required number of tags (more exactly tag-text pairs).
In our example, suppose the document contains just 2 {\it keywords} as opposed to the required 9, we would not be able to sample according to the original distribution. In order to address this problem, we propose re-distributing the remaining ``tag-budget'' to the other tag types in the summary structure, by repeated sampling and re-normalization of the importance distribution on tags, up to the desired summary size.

An example is shown in Table \ref{tab:rounds}. Let the desired summary size $|S| = 30$. In Round 1, the initial summary size $|S| = 0$. In step 1.1, multiplying the desired summary size with the probability of \Tag{actor} gives us 15 instances of this tag to be added into the summary. Since the number of actors in the document is 30, we can include the top 15 actors into the summary. However, in step 1.2, the number of \Tag{keyword} tags to be included turns out to be 9, while we have only 2 keyword tags in the document. We include both keywords into the summary and note that a probability mass of 0.3 is available for redistribution. Continuing in step 1.3, we include 6 \Tag{trivia} into the summary. Since we still require 7 more tags, we need to continue to sample from the top. However, before we do so, we redistribute the probability mass of \Tag{keyword} (0.3) in proportion to the remaining available tags. In this case, \Tag{actor}, the top-ranked tag is still available for inclusion as is the bottom-ranked tag \Tag{trivia}. Hence, the probability is distributed in proportion to their existing probability mass before round 2. In round 2, we again start with \Tag{actor} in step 2.1. Repeating the calculations as in round 1, we end up with 5 additional \Tag{actor} and 2 \Tag{trivia} tags. Overall, the final summary consists of 20 \Tag{actor}, 2 \Tag{keyword} and 8 \Tag{trivia} tags. Once the tags have been chosen, they need to be filled with the appropriate text values. The top-ranked text values are preferred except when a co-occurring set of tags need to be populated as described in the previous section.

\begin{table}[ht]
\small
\begin{center}
\begin{tabular}{|c|c|c|c|c|c|}
\hline
{\bf Step}&{\bf Tag} & {\bf Prob.} & \#tags  & \#tags  & \#tags\\
          &          &             & rema-    &  to be      & actually \\
          &          &             & ining    &  added to    & added \\
          &          &             &            &  $S$         & (total \#tags\\
          &          &             &            &            & in $S$\\
\hline
\hline
\multicolumn{6}{|c|}{{\bf Round 1:} $|S| = 0$}\\
\hline
1.1 & \Tag{actor}   & 0.5       &   30            & 15         & 15 (15)\\
\hline
1.2 & \Tag{keyword} & 0.3       &   2             & 9          & 2 (2)\\
\hline
1.3 & \Tag{trivia}  & 0.2       &    15           & 6          & 6 (6)\\
\hline 
\multicolumn{6}{|c|}{{\bf Round 2:} $|S| = 23$}\\
\hline
2.1 & \Tag{actor}   & $5/7\approx0.7$       &   15            & 5        & 5 (20) \\
\hline
-  & \Tag{keyword} & 0       &     0             & 0        & 0 (2)\\
\hline
2.2 & \Tag{trivia}  & $2/7\approx0.3$       &   9            & 2        & 2 (8) \\
\hline 
\multicolumn{6}{|c|}{$|S| = 30$}\\
\hline
\end{tabular}
\caption{Generating the summary with $|S| = 30$}
\label{tab:rounds}
\end{center}
\end{table}
\normalsize
\chapter{Experiments}
\label{sec:6expts}

The goal of our experiments was to determine how good our techniques
are in generating summaries of various sizes. We describe our datasets
and metrics in more detail in the following.

\section{Datasets}
We used two datasets for both set of experiments which are summarized
in Table \ref{tab:datasets}. Both datasets -- Movie and People -- were
extracted from the IMDB corpus (available from
\url{http://www.imdb.com}). Out of the corpus of available documents,
8 documents from each dataset were chosen for summarization. The list
of these documents for each dataset is shown in Table \ref{tab:test}.

\begin{table}[ht]
\small
\begin{center}
\begin{tabular}{|c|c|c|c|}
\hline
{\bf Dataset} & {\bf \#files} & {\bf Example} & {\bf \#tags}\\
              & {\bf (corpus)}& {\bf tags}    & {\bf (unique)} \\
\hline
Movie  & 200,000 & \Tag{title},\Tag{director}           &   39 \\
       &         & \Tag{actor}, \Tag{role} & \\
       &         & \Tag{goofs}, \Tag{alt\_versions} & \\
\hline
People & 150,000 & \Tag{birthdate}, \Tag{spouse}        & 11 \\
       &         & \Tag{acted\_in}, \Tag{biography} & \\
\hline
\end{tabular}
\caption{Description of Datasets}
\label{tab:datasets}
\end{center}
\end{table}

\normalsize
\begin{table}[ht]
\small
\begin{center}
\begin{tabular}{|c|c|c|}
\hline
{\bf Dataset} & {\bf Filename} & {\bf \#tags}\\
              &                & {\bf total}\\
\hline
Movie & American Beauty & 832  \\
      & Ocean's Eleven & 795 \\
      & Kill Bill Part II & 153 \\
      & Saving Private Ryan & 1121 \\
      & The Last Samurai & 429 \\
      & The Usual Suspects & 617 \\
      & Titanic & 1681 \\
      & 2001: A Space Odyssey & 1107 \\
\hline
\hline
People & Matt Damon & 116 \\
       & Ben Affleck & 136 \\
       & Tom Cruise & 150 \\
       & Leonardo DiCaprio & 79 \\
\hline
\end{tabular}
\caption{Documents used for Summarization}
\label{tab:test}
\end{center}
\end{table}
\vspace*{-0.1in}

\normalsize
\section{Metrics}
We conducted an intrinsic evaluation of summaries of various
sizes. That is, evaluators were asked to judge a summary in and of
itself (see \cite{Mani:NTCIR2001} for a more detailed explanation of
intrinsic evaluations). The evaluators were asked to provide a grade
to the summary ranging from 1 to 7 (1 -- extremely bad, 2 -- pretty
bad, 3 -- bad, 4 -- ok, 5 -- good, 6 -- pretty good, 7 -- perfect)
based on whether the summary reflected its source. They were asked to
consider the importance of the tag-text pairs selected, the coverage
offered, and to also take into account the hard restriction of summary
size. In effect, they were asked to grade the summary based on how
well it made use of the space available.

We used a total of 6 evaluators to conduct the evaluations and each
summary had at least 3 evaluations by 3 separate evaluators (the setup
is described in more detail in the next section). Once they submitted
their evaluations, we conducted a short survey to understand how they
arrived at their grade and their general impressions. We report on
both the grades of the evaluators as well as their impressions and
infer some trends.

\section{Evaluation Setup}

\begin{table}[ht]
\small
\begin{center}
\begin{tabular}{|c|c|c|}
\hline
              & {\bf Size} & {$\alpha$} \\
 \hline
              &  5         & 1, 0.8\\
{\bf Movie}   &  10         & 1, 0.8, 0.6\\
              &  20         & 1, 0.8, 0.6\\
\hline
\hline
{\bf People} &  5         & 1, 0.6\\
             &  10         & 1, 0.6\\
\hline
{\bf TOTAL} & \multicolumn{2}{c|}{64+16 = 80}\\
\hline
\end{tabular}
\caption{Summaries used for evaluation.}
\label{tab:setup}
\end{center}
\end{table}

\vspace*{0.2in}
\normalsize
\paragraph*{Automatically Generated Summaries}
The set of summaries generated for evaluation is tabulated in Table
\ref{tab:setup}. A total of 8 summaries per document were generated
for the movie documents while a total of 4 summaries per document were
generated for the people documents. We experimented with 3 values of
the parameter $\alpha$ for choosing tags, from $1.0$ (typicality
only), $0.8$ ($0.8 P_{typ} + 0.2 P_{spe}$) and $0.6$ ($0.6 P_{typ} +
0.4 P_{spe}$).

We did not generate a 5-element summary with $\alpha = 0.6$ for the
movie documents since this value of $\alpha$ eliminated the
intuitively most important tag, \Tag{title} from the summary, which
made it not worth evaluating. For the people documents, we did not
generate summaries for $\alpha = 0.8$ since the same summary was
generated as for $\alpha = 1.0$.

For text selection, we chose approximately the same parameter values
$\lambda = 0.49$ and $\mu = 0.48$ for short text and entities. This
was in effect giving the tag and document context approximately the
same importance, while the corpus context was used as a last resort
(often to resolve importance in the case of ties between two
values). For long text, we used only the centroid query method to
generate text values (recall that for long text, we only take the tag
context into consideration).

\paragraph*{Human-generated Summaries} An important thing that we
realized early on is that there is no single summary which can be
considered the best. In fact, there have been studies conducted for
text summarization which show that even human-generated summaries may
not agree on a majority of the content and that the same person
generating a summary at two different times may not generate the same
summary \cite{Mani:NTCIR2001, Jing:AAAI1998}. Therefore, while we
instructed the evaluators that more than one summary of the same size
could have the same grade, we also added an additional
\emph{human-generated summary} to each evaluation batch. The
human-generated summaries were constructed by people who did not
participate in the final evaluation. Additionally, the evaluators were
not told that human-generated summaries were included in the
evaluation and so, would not be biased by its presence. We were able
to gain additional understanding of the potential of our techniques
with human-generated summaries as a basis for comparison.

In total, 112 summaries were evaluated by 6 different evaluators. Each
summary was evaluated by at least 3 evaluators. Among the 112
summaries, 80 were generated by our techniques and 32 were
human-generated summaries. A total of 336 evaluations were collected
(264 for the Movie dataset and 72 from the People dataset).

\section{Results}

\begin{table*}[ht]
\small
\begin{center}
\begin{tabular}{|c|c|c|c|c|l|}
\hline
{\bf Dataset} & {\bf Size} & \multicolumn{3}{|c|}{$\alpha$ values} & \\
\hline
\hline
      &    & 1.0           & 0.8            & 0.6          & {\bf Total}\\
      &    &               &                &              & (across $\alpha$)\\
\hline
Movie & 5  & 8/8 (100\%)   & 5/8 (62.5\%)   & --            & {\bf 13/16 (81.25\%)}\\
      & 10 & 8/8 (100\%)   & 7/8 (87.5\%)   & 1/8 (12.5\%) & {\bf 16/24 (66.6\%)}  \\
      & 20 & 7/8 (87.5\%)  & 7/8 (87.5\%)   & 4/8 (50\%)   & {\bf 18/24 (75\%)}    \\
\hline
&{\bf Total}  &  {\bf 23/24 (95.8\%)}& {\bf 19/24 (79.1\%)}& {\bf 5/16 (31.2\%)}& {\bf 47/64 (73.4\%)}\\
&(across sizes)&&&&\\
\hline
\hline
People & 5  & 3/4 (75\%) &      --          &  1/4 (62.5\%)  & {\bf 4/8 (50\%)}\\ 
       & 10 & 4/4 (100\%) &     --           & 4/4 (100\%)   & {\bf 8/8 (100\%)}\\
\hline
&{\bf Total} & {\bf 7/8 (87.5\%)} &     --          & {\bf 5/8 (62.5\%)}  & {\bf 12/16 (75\%)}\\
&(across sizes)&&&&\\
\hline
\end{tabular}
\caption{Tabulation of average and above average grades (4,5,6,7) across all documents. Grades are reported only if at least 2 evaluators agreed on it.}
\label{tab:all}
\end{center}
\end{table*}

\normalsize
In the following, we report on the grades provided by the evaluators
for various classes of summaries only if the relevant grade has been
provided by at least 2 out of 3 evaluators. For specific examples of
grades, please refer to Tables \ref{tab:best1} and \ref{tab:best2}, which tabulate the XML documents and the best, worst and as-good-as-human-generated
summaries. For examples of summaries generated by our system, please
refer to the Appendix.

Table \ref{tab:all} presents the summary of our results. It shows the
number of summaries graded average and above average for various
values of $\alpha$ and different summary sizes. Each cell contains an
entry of the form x/y (z\%), where y is the total number summaries in
that category (for example, 8 is the number 10-element summaries with
$\alpha = 1.0$, in the Movie dataset), x is the number summaries in
that category which were graded average and above by at least 2
evaluators, while z shows the percentage. The ''Totals'' rows and
column provide aggregated numbers across all $\alpha$ values and
across all sizes for both datasets.

\subsection{Analysis}

\paragraph*{\bf Impact of $\alpha$} It is clear from Table
\ref{tab:all}, that while $\alpha = 1, 0.8$ result in good summaries,
an $\alpha = 0.6$ value results in low quality summaries (especially
for the Movie dataset). For the Movie dataset, the total (across all
$\alpha$), despite being well over the 50\% mark, suffers because of
the low grades for summaries with $\alpha = 0.6$. If we eliminate
these low quality summaries from our computation (for the Movie
dataset only), we find that a total of 15/16 (93.7\%) of 10-element
summaries and 14/16 (87.5\%) of 20-element summaries score average and
above average grades. For the People dataset, a lower value of
$\alpha$ slightly reduces the effectiveness of the summaries. In
total, 5/8 (62.5\%) summaries were given average and above average
grades for this dataset.

We thus conclude that our evaluators preferred highly typical tags for
the given datasets. In order to understand why specialty was not
playing a bigger role in the grading process, we questioned the
evaluators. Many evaluators expressed the opinion that, while it was
nice to see a special tag (in the Movie dataset, \Tag{trivia},
\Tag{goof}, had low typicality and high specialty), they did not need
to see more than one or two of them. Whenever they felt that there
were too many special tags, those summaries were ranked lower. We
concluded from this that it was not specifically the specialty
component of our model that was at fault, but the special tags in our
current datasets. These tags were not all that appealing to the
evaluators. For example, an \Tag{oscar\_winner} tag, occurring
multiple times may have been much more appealing than \Tag{trivia},
but we did not have such a tag in our dataset.

\paragraph*{\bf Impact of summary size} It is clear from Table
\ref{tab:all} that the best $\alpha$ (in our case $\alpha = 1.0$ for
both datasets) results in consistently good summaries across \emph{all
sizes} (95.8\% of summaries for Movie dataset and 87.5\% of summaries
for People dataset). This is an important point favoring our
techniques. The larger the summary, the larger the options to choose
from and the larger the chance of junk being selected.  The consistent
good grades across the different sizes shows that our techniques
succeed in choosing the right elements for inclusion in the summary as
the desired size increases.

\paragraph*{\bf Impact of text values} Our summaries contained both
long as well as short text values and entities. In order to understand
how the choice of text values impacted the grades, we again questioned
the evaluators. For long text values such as \Tag{trivia}, \Tag{plot}
and \Tag{goof}, they were not particularly interested in the exact
value chosen, but were happy to see that they were present in the
summary. Evaluators who hand-generated the summaries also had a
similar opinion -- that they didn't really see a good criterion to
choose one text value over another, and that any of them would be
acceptable.

For the short text values and entities, which are more easily
readable, higher importance was given. For values such as those for
\Tag{actor} (in Movies) and \Tag{acts\_in} (in People), it was
mandatory that the most important values (lead actors, famous movies)
be chosen. In the case of less typical tags, such as \Tag{keyword},
only the relevance of the keyword to the movie was taken into account,
rather than the best keyword among those available in the source.

Hence, we conclude that it is extremely important to have robust
techniques to choose the best values for entities and short text. And
our techniques seem to work well for entities and short text. For
longer text, it does not seem to be all that important. However,
retaining the flexibility to generate diverse values may be essential
for other datasets.

\begin{table*}[ht]
\small
\begin{center}
\begin{tabular}{|l|l||l|l|}
\hline
\multicolumn{2}{|c|}{Best (grades 6,7)} & \multicolumn{2}{|c|}{Worst (grades 1--3)}\\
\hline
{\bf File} & {\bf Size} &{\bf File} & {\bf Size}\\
					 & ($\alpha = 1$) & & ($\alpha = 0.6$)\\
           &                & & (in most cases)\\
\hline
American Beauty & 10 & American Beauty &  10, 20\\
Kill Bill - 2 & 5 & Kill Bill - 2 & 5, 10, 20\\
Ocean's Eleven & 10 & Ocean's Eleven & 5, 10, 20\\
Saving Private Ryan & 5, 10, 20 & Saving Private Ryan & 10\\
The Last Samurai & 5, 10, 20 & 2001: A Space Odyssey & 10, 20\\
Titanic & 10 & Titanic & 5, 10\\
Usual Suspects & 5, 10 & Usual Suspects & 10\\
Cruise & 5 & &\\
\hline
\end{tabular}
\caption{Best and Worst Summaries.}
\label{tab:best1}
\end{center}
\end{table*}
\normalsize

\begin{table*}[ht]
\small
\begin{center}
\begin{tabular}{|l|l|}
\hline
\multicolumn{2}{|c|}{As-good-as human-generated}\\
\hline
{\bf File} & {\bf Size}\\
					 & ($\alpha = 1$)\\
\hline
Kill Bill - 2 & {\bf 5}\\
Saving Private Ryan & {\bf 5, 10, 20}\\
The Last Samurai & {\bf 5, 10, 20}\\
Usual Suspects & {\bf 10, 20}\\
American Beauty & {\bf 20}\\
\hline
\end{tabular}
\caption{As-good-as (human-generated) Summaries.}
\label{tab:best2}
\end{center}
\end{table*}
\normalsize
\chapter{Conclusions and Future Work}
\label{sec:7conc}
Our focus in this paper was to provide general-purpose techniques to generate
concise, generic summaries \emph{automatically} for a given XML document. We proposed a framework and model for ranking tags and text. We described an algorithm for generating size-constrained summaries. Finally, we showed through a user study that our techniques are able to generate good summaries for a range of different summary sizes and made recommendations on how to set the tuning parameters.

\noindent
There are at least a couple of directions for future work. First, many evaluators were of the opinion that the text values were sometimes too long (for tags such as \Tag{plot}, \Tag{trivia}, etc.). One direction of future work is to use text summarizers to shorten these values. We experimented with this in our previous work \cite{Ramanath:dbrank2008}, but there is a need for a more comprehensive model for rewriting both structure as well as text. Second, it would be interesting to develop ranking functions for different kinds of text. In this work, we only considered entities and regular text. In addition, we may also consider numbers and special methods for ranking them.

\small
\bibliographystyle{plain}
\bibliography{paper}

\normalsize
\appendix

\addappheadtotoc

\chapter{}
\label{appendix:a}
\small
All evaluated summaries and grade tabulation are available from\\ {\tt http://mpi-inf.mpg.de/$^\sim$ramanath/Summarization}.

\noindent
Example summaries for \emph{The Last Samurai} and \emph{The Usual Suspects} ($\alpha = 1.0$, 5-element, 10-element).

\scriptsize
\noindent
$<$movie$>$ \\
\hspace*{0.1in}  $<$title$>$ \emph{Last Samurai, The} $<$/title$>$ \\
\hspace*{0.1in}  $<$prod\_year$>$ \emph{2003} $<$/prod\_year$>$ \\
\hspace*{0.1in} $<$director$>$ \emph{Zwick, Edward} $<$/director$>$ \\
\hspace*{0.1in} $<$colorinfo$>$ \emph{Color} $<$/colorinfo$>$ \\
\hspace*{0.1in}  $<$cast$>$$<$casting$>$ \\
\hspace*{0.2in}    $<$actor$>$ \emph{Cruise, Tom} $<$/actor$>$ \\
\hspace*{0.1in}  $<$/casting$>$$<$/cast$>$ \\
$<$/movie$>$ \\

\scriptsize
\noindent
$<$movie$>$ \\
\hspace*{0.1in}  $<$title$>$	\emph{Usual Suspects, The} $<$/title$>$ \\
\hspace*{0.1in}  $<$prod\_year$>$	\emph{1995} $<$/prod\_year$>$ \\
\hspace*{0.1in}  $<$prod\_lang$>$	\emph{English} $<$/prod\_lang$>$ \\
\hspace*{0.1in}  $<$director$>$	\emph{Singer, Bryan} $<$/director$>$ \\
\hspace*{0.1in}  $<$genres$>$ \\
\hspace*{0.2in}    $<$genre$>$ \emph{Crime} $<$/genre$>$ \\
\hspace*{0.2in}    $<$genre$>$	\emph{Thriller} $<$/genre$>$ \\
\hspace*{0.1in}  $<$/genres$>$ \\
\hspace*{0.1in}  $<$colourinfo$>$ \emph{Color (Technicolor)} $<$/colourinfo$>$ \\
\hspace*{0.1in}  $<$cast$>$ \\
\hspace*{0.2in}    $<$casting$>$ \\
\hspace*{0.3in}      $<$actor$>$	\emph{Spacey, Kevin} $<$/actor$>$ \\
\hspace*{0.3in}      $<$role$>$	\emph{Roger'Verbal'Kint} $<$/role$>$ \\
\hspace*{0.2in}    $<$/casting$>$ \\
\hspace*{0.2in}    $<$casting$>$ \\
\hspace*{0.3in}      $<$actor$>$	\emph{Byrne, Gabriel} $<$/actor$>$ \\
\hspace*{0.2in}    $<$/casting$>$ \\
\hspace*{0.1in}  $<$/cast$>$ \\
$<$/movie$>$\\

\noindent
\small
The summary for \emph{Benjamin Affleck} ($\alpha = 0.8$, 10-element).

\scriptsize
\noindent
$<$person$>$\\
\hspace*{0.1in}$<$name$>$ Ben Affleck $<$/name$>$\\
\hspace*{0.1in}$<$produced$>$\\
\hspace*{0.2in}$<$movie$>$ Crossing Cords $<$/movie$>$\\
\hspace*{0.1in}$<$/produced$>$\\
\hspace*{0.1in}$<$acts\_in$>$ \\
\hspace*{0.2in}$<$movie$>$ third wheel, the $<$/movie$>$\\
\hspace*{0.2in}$<$role$>$ Michael $<$/role$>$\\
\hspace*{0.1in}$<$/acts\_in$>$\\
\hspace*{0.1in}$<$acts\_in$>$ \\
\hspace*{0.2in}$<$movie$>$ good will hunting $<$/movie$>$\\
\hspace*{0.2in}$<$role$>$ Chuckie Sullivan $<$/role$>$\\
\hspace*{0.1in}$<$/acts\_in$>$\\
\hspace*{0.1in}$<$acts\_in$>$$<$movie$>$ voyage of the mimi, the $<$/movie$>$\\
\hspace*{0.1in}$<$/acts\_in$>$ \\
\hspace*{0.1in}$<$acts\_in$>$$<$movie$>$ pearl harbor $<$/movie$>$\\
\hspace*{0.1in}$<$/acts\_in$>$ \\
\hspace*{0.1in}$<$biography$>$ \\
\hspace*{0.2in}$<$author$>$ trendekid at aol.com $<$/author$>$\\
\hspace*{0.2in}$<$text$>$ benjamin geza affleck was \ldots $<$/text$>$\\
\hspace*{0.1in}$<$/acts\_in$>$\\
$<$/person$>$\\

\scriptsize
\normalsize

\end{document}